\documentclass{aipproc}

\newcommand{\be}{\begin{equation}}
\newcommand{\ee}{\end{equation}}
\newcommand{\bea}{\begin{eqnarray}}
\newcommand{\eea}{\end{eqnarray}}

\layoutstyle{6x9}

\SetInternalRegister\hbadness{8000} % pseudo latin isn't breaking very well :-)

\newcommand\doingARLO[2][]{%
  \ifx\mmref\undefined #1\else #2\fi
}

\begin{document}
\def\esp #1{e^{\displaystyle{#1}}}
\def\slash#1{\setbox0=\hbox{$#1$}#1\hskip-\wd0\dimen0=5pt\advance
       \dimen0 by-\ht0\advance\dimen0 by\dp0\lower0.5\dimen0\hbox
         to\wd0{\hss\sl/\/\hss}}\def\ink {\int~{d^4k\over (2\pi)^4}~}

\title
      [QCD Supersymmetry and Low Scale Gravity]
      {QCD Supersymmetry and Low Scale Gravity}

%\classification{43.35.Ei, 78.60.Mq}
%\keywords{Document processing, Class file writing, \LaTeXe{}}

\author{Alessandro Cafarella$^*$ Claudio Corian\`o}{
  address={Dipartimento di Fisica dell' Universita' di Lecce and INFN-Lecce, 
Via Arnesano, 73100 Lecce, Italy.
E-mail: alessandro.cafarella@le.infn.it, claudio.coriano@le.infn.it},
  email={alessandro.cafarella@le.infn.it, Claudio.coriano@le.infn.it},
  %thanks={This work was commissioned by the AIP}
}\footnote{Presented by C. Corian\`o at the VI Conference on Quark Confinement and the Hadron Spectrum, Villasimius, Sardinia, Italy, 21-25 September 2004}

\author{T. N. Tomaras}{
  address={Department of Physics and Institute for Plasma Physics,
University of Crete and FORTH, Heraklion, Crete, Greece
E-mail: tomaras@physics.uoc.gr},
  email={alessandro.cafarella@le.infn.it, claudio.coriano@le.infn.it},
  %thanks={This work was commissioned by the AIP}
}

%\begin{abstract}

%\end{abstract}

\date{\today}

\maketitle

Theories with large extra dimensions \cite{ED} invoke a brane picture 
of the universe, with matter confined on a brane embedded 
in a higher D-dimensional space $(D=4 + N)$, and only gravity 
free to propagate in the extra dimensions. 
A certain number, say $n$, of the $N$ extra dimensions may be large,
with size of the order of a millimeter. These scenarios are characterized 
by a low fundamental scale for gravity, $M_*$, related to the Planck scale $M_{Pl}$
by $M_{Pl}^2= M_*^{n+2}V_{(n)}$, with $V_{(n)}$ the volume of the extra dimensions.
For $n=2$, $M_*$ can be of the order of a TeV if the typical size of the large 
extra dimensions is a millimeter. At the LHC, for QCD factorization scales above $M_{*}$, 
gravity becomes strong and hadronic collisions should be characterized 
by a rich new phenomenology. In particular, mini black holes of mass 
$M_{BH}\approx M_{*}$ are expected to be produced copiously \cite{DL} 
(see \cite{HSU} for a discussion of some quantum aspects).

Mini black holes are hot, characterized by a temperature  
which is inversely proportional to their mass $M_{BH}$. Their formation 
takes place in (parton-parton) collisions for impact parameters of the 
order of the size of the horizon ($r_H$) 
\begin{equation}
r_H= \frac{1}{\sqrt{\pi}M_*}\left(\frac{M_{BH}}{M_*}\right)^
{\frac{1}{n+1}}\left(\frac{8\Gamma\left(\frac{n+3}{2}\right)}{n+2}\right)
^{\frac{1}{n+1}}\,.
\label{horizon}
\end{equation} 
corresponding to the collision energy $E\sim M_{BH}$ in the center of mass frame.
For $n>0$ the relation 
between $r_H$ and $M_{BH}$ becomes nonlinear and the presence of $M_*$ in 
the denominator of 
Eq.~(\ref{horizon}) in place of $M_{Pl}$ increases the size of the horizon 
for a given $M_{BH}$. For $M_{BH}/M_*\sim 5$ and $M_*=1$ TeV the size of the 
horizon is around $10^{-4}$ fm and decreases with increasing $n$.
A good approximation to the partonic cross section for producing a mini black hole is 
$\sigma_{BH}\approx \pi r_H^2$, 
the geometrical one. It can be folded with parton distributions $(f(x, Q^2))$ to give predictions, for instance, 
for total cross sections 
\begin{equation}
\sigma(pp \rightarrow BH+X) =\frac{1}{ s}
\sum_{a,b}\int^{{s}}_{M^2_{BH,min}}dM_{BH}^2 \nonumber\\
 \times  \int^1_{x_{1,min}} 
 {dx_1\over x_1}
 f_a(x_1,Q^2)\sigma_{BH}f_b(x_2,Q^2),
\label{cross}
\end{equation}
where $x_1$ and $x_2={M_{BH}^2/( x_1 s)}$ are the momentum fractions of the 
initial partons and $x_{1,min}=M^2_{BH}/s$. The factorization scale $Q$ 
is of the order of $1/r_H$. The absorption/emission cross section depends 
sensitively on the greybody factors of the black hole, 
which are energy dependent. The choice of either constant 
or full energy-dependent greybody factors, which are known for static 
(Schwarzschild) mini black hole solutions \cite{PKanti}, gives widely 
different results \cite{Sarcevic}. 
The known analytical expressions of the greybody factors at low frequencies 
are of limited help in the prediction of the event rates at the LHC,
but these can be computed numerically \cite{HK}. Of particular relevance 
would be the numerical study of the greybody factors for 
Kerr solutions, since black holes, in general, will be produced with 
non-vanishing angular momentum.

Studies of the $p_T$ distributions 
show a much larger signal 
compared to the fast falling QCD background \cite{Sarcevic}, even for $M_*$ as high as 5 TeV, 
starting at $P_T\sim 50-200$ GeV and up. The dependence on the number of extra dimensions $n$ is also significant.
A second sensitivity in the prediction of event rates 
comes from the integration over the invariant mass $M_{BH}$ for $M_{BH}$ close to $M_{Pl}$, 
since the semiclassical picture of the formation and decay of the black hole is not valid any longer. 
In all the studies presented so far larger multiplicites of the final states and broader $p_T$ 
distributions appear to be a striking signature of mini black hole formation in hadron collisions.
In the most optimistic scenario in which both low energy gravity and supersymmetry will be discovered at the LHC, then the multiplicities of the final state in the decay of the black hole should 
grow even faster from what inferred from these studies. However, it is important to keep in mind that a part 
of the energy available in the collision is loss into gravitational emission, and only 
a fraction of it remains available for the hadronization, which would imply reduced multiplicities. 

The time scales for the black hole decay into partons and the QCD hadronization scale 
are largely separated and the decay of the black hole is, essentially, 
instantaneous. 
Hadronization takes place soon after the partons, which are emitted in an approximate s-wave, cross the horizon. 
The emissions of single partons are assumed to be uncorrelated, and can be described by a 
multinomial distribution, while the hadronization is studied either using 
Monte Carlo \cite{Webber} or renormalization group equations \cite{CCT}. 

The computation of the cumulative probabilities to produce any number (K) of hadrons of type 
$h$ by the decay of the black hole are obtained from the multinomial 
distribution multiplied by the fragmentation probabilities of each elementary 
state to $h$ and summing over all possible emissions \cite{CCT}
\begin{equation}
\textrm{Pr}_{\textrm{cum h}}(K,Q)\equiv \sum_{n_f, n_i} {K!\over \prod_f n_f! \prod_i n_i!}\prod_f 
\left(p_f <D_f^h(Q_F)>\right)^{n_f}\prod_i \left(p_i <D_i^h(Q_F)>\right)^{n_i},
\label{probab}
\end{equation}
where $i$ is summed over gluons, photons and a set or remainder states, f runs over the 
quark flavours, while $K= n_f + n_i$.
In (\ref{probab}) the $<D_{i,f}^h(Q_F)>$ are the first moments of the fragmentation functions of a 
parton/photon $k$ to a hadron $h$ at a scale $Q_F$. 
The sum is over all the main hadronic states. The fragmentation scale $Q_F$ is related to the number 
of fundamental decaying states $N_m$ to which the black hole couples in a democratic way and to its mass $M_{BH}$ by 
$Q_f= M_{BH}/N_m$, where our knowledge 
of the multiplicity $N_m$ is clearly approximate. Obtaining a good estimate of $N_m$
is important for studies of the multi-jet structure of the events at the LHC, but is less relevant for 
cosmic ray studies. In this latter case the evolution of the air shower after the decay of the black hole 
washes out the information on small variations in the original multiplicities in the decay. 
At this time, the only known formulas available for $N_m$ come from a semiclassical analysis. 
We recall that in cosmic ray physics mini black hole events can be triggered 
by neutrinos scattering off nucleons in the atmosphere. An analysis of the lateral distributions of showers and of the corresponding multiplicities shows that intermediate mini black hole resonances 
are respectively much wider and larger compared to ordinary air showers \cite{CCT}, 
in agreement with the fireball picture of the decay which has emerged from LHC studies. 

Proposals for the best approximation to $N_m$ are several. 
In \cite{DL} was suggested to use 
\begin{equation}
    N_m  = \frac{2\pi}{n+1}
    \left(\frac{M_{BH}}{M_*}\right)^\frac{n+2}{n+1}\,\left(\frac{8\Gamma\left(\frac{n+3}{2}\right)}{n+2}\right)^{\frac{1}{n+1}}\frac{1}{\sqrt{\pi}},
\label{nav}
\end{equation}
but there are variants of it. Other expressions include a correction factor $\rho$ coming 
from a more detailed analysis of the Hawking formula for the semiclassical decay 
which takes into account the corresponding greybody $(\Gamma_s)$ factors more accurately \cite{Cavaglia1}. Then $N_m=\rho S_0$ 
with $S_0$ being the entropy of the black hole
and 
\begin{equation}
\rho=\frac{ \sum_s c_s\, f_s\, {\Gamma}_s\, \Gamma(3)\, \zeta(3)}
{ \sum_s c_s \,{f'}_s\, {\Gamma}_s\, \Gamma(4)\, \zeta(4)},
\end{equation} 
which is expressed in terms of the greybody factors and certain numerical coefficients $(c_s, f_s, {f'}_s)$ 
dependent on the spin $s$ of the fields propagating over the black hole background. 
As we have already mentioned, the issue of gravitational energy emission during the formation of the 
black hole and during its decay remains open. Work in this direction can follow closely 
some of the recent results on the study of quasi-normal modes for ordinary black holes in 4 dimensions 
aimed at the detection of gravitational waves \cite{Cardoso}.

%\newpage


\begin{thebibliography}{99}
\bibitem{ED}
N. Arkani-Hamed, S. Dimopoulos and G. Dvali,
Phys. Lett. {\bf B429}, 263 (1998);
I. Antoniadis, N. Arkani-Hamed, S. Dimopoulos and G. Dvali,
Phys. Lett. {\bf B436}, 257 (1998);
L.~Randall and R.~Sundrum, Phys. Rev. Lett. {\bf 83}, 3370 (1999);
L.~Randall and R.~Sundrum, Phys. Rev. Lett. {\bf 83}, 4690 (1999).
\bibitem{DL}S.~Dimopoulos and G.~Landsberg, Phys.~Rev.~Lett. \textbf{87},
161602 (2001).
\bibitem{HSU} S.D.H. Hsu, Phys.Lett.{\bf B555} 92, (2003);
V. P. Frolov, D. Stojkovic, Phys.Rev.{\bf D 67} 084004, (2003). 
\bibitem{PKanti} P. Kanti and J. March-Russell, Phys. Rev.{\bf D 67}, 104019 (2003).
\bibitem{Sarcevic} I. Mocioiu, Y. Nara and I. Sarcevic, Phys. Lett. B {\bf 557}, 87 (2003).
\bibitem{HK} P. Kanti and C.M. Harris, JHEP {\bf 0310} 014,(2003). 
\bibitem{Webber} C.M. Harris, P. Richardson and B.R. Webber, JHEP 0308,033, (2003).
\bibitem{CCT} A. Cafarella, C. Corian\`o and T.N. Tomaras, [arXiv:hep-ph/0410358], [arXiv:hep-ph/0410190]. 
\bibitem{Cavaglia1} M. Cavagli\`a Phys.Lett. {\bf B569}, 7, (2003), 
M. Cavagli\`a and S. Das, Class.Quant.Grav. {\bf 21}, 4511, (2004);
\bibitem{Cardoso} V. Cardoso, O. J.C. Dias, J.P.S. Lemos, Phys. Rev. {\bf D 67}, 064026 (2003); 
K.D. Kokkotas, Living Rev.Rel.{\bf 2}, 2, (1999); see also  
H. Kodama and A. Ishibashi, Prog.Theor.Phys.{\bf 110} 701, (2003). 

\end{thebibliography}
\end{document}